\documentclass[pdflatex,sn-basic]{sn-jnl}%

\usepackage{graphicx}%
\usepackage{multirow}%
\usepackage{amsmath,amssymb,amsfonts}%
\usepackage{amsthm}%
\usepackage{subfigure}%
\usepackage{mathrsfs}%
\usepackage[title]{appendix}%
\usepackage{xcolor}%
\usepackage{textcomp}%
\usepackage{manyfoot}%
\usepackage{booktabs}%
\usepackage{algorithm}%
\usepackage{algorithmicx}%
\usepackage{algpseudocode}%
\usepackage{listings}%
\usepackage{bm}
\newcommand{\matt}[1]{\mathbf{#1}}
\newcommand{\vecc}[1]{\mathbf{#1}}
\newcommand{\norm}[1]{\left\lVert#1\right\rVert}
\newcommand*{\defeq}{\mathrel{\vcenter{\baselineskip0.5ex \lineskiplimit0pt
                     \hbox{\scriptsize.}\hbox{\scriptsize.}}}%
                     =}

\usepackage{cleveref}

\crefname{equation}{}{}
\usepackage{tikz}
\usepackage{afterpage}
\usepackage{pgfplots}
\pgfplotsset{compat=1.3}
\usepgfplotslibrary{groupplots}
\pgfkeys{/pgf/number format/.cd,1000 sep={}}
\usepackage{algorithm}
\usepackage{algpseudocode}
\algnewcommand\And{\textbf{and} }

\usepackage{siunitx}
\DeclareSIUnit{\deg}{deg}
\usepackage[inline]{enumitem}
\usepackage{caption}

\theoremstyle{thmstyleone}%

\theoremstyle{thmstyletwo}%

\theoremstyle{thmstylethree}%

\begin{document}

\title[Initial Trajectory Assessment of the RAMSES Mission to (99942) Apophis]{Initial Trajectory Assessment of the RAMSES Mission to (99942) Apophis}

\author[1]{\fnm{Andrea C.} \sur{Morelli}}\email{andreacarlo.morelli@polimi.it}
\equalcont{These authors contributed equally to this work.}

\author[1]{\fnm{Alessandra} \sur{Mannocchi}}\email{alessandra.mannocchi@polimi.it}
\equalcont{These authors contributed equally to this work.}

\author*[1]{\fnm{Carmine} \sur{Giordano}}\email{carmine.giordano@polimi.it}

\author[1]{\fnm{Fabio} \sur{Ferrari}}\email{fabio1.ferrari@polimi.it}

\author[1]{\fnm{Francesco} \sur{Topputo}}\email{francesco.topputo@polimi.it}

\affil[1]{\orgdiv{Department of Aerospace Science and Technology}, \orgname{Politecnico di Milano}, \orgaddress{\street{via La Masa, 34}, \city{Milan}, \postcode{20156}, \country{Italy}}}

\abstract{(99942) Apophis is a potentially hazardous asteroid that will closely approach the Earth on April 13, 2029. Although the likelihood of an impact has been ruled out, this close encounter represents a unique opportunity for planetary science and defense. By investigating the physical and dynamical changes induced by this interaction, valuable insights into asteroid cohesion, strength, and internal structure can be obtained.  In light of these circumstances, a fast mission to Apophis holds great scientific importance and potential for understanding potentially hazardous asteroids. To this aim, ESA proposed the mission RAMSES (Rapid Apophis Mission for SEcurity and Safety) to reach Apophis before its close encounter. In this context, the paper focuses on the reachability analysis of (99942) Apophis, examining thousands of trajectories departing from Earth and reaching the asteroid before the fly-by, using a low-thrust spacecraft. A two-layer approach combining direct sequential convex programming and an indirect method is employed for fast and reliable trajectory optimization. The results reveal multiple feasible launch windows and provide essential information for mission planning and system design.}


\maketitle

\section{Introduction}
(99942) Apophis is a potentially hazardous asteroid with a diameter of about 370 metres that caused a brief period of concern in December 2004 when initial observations indicated a probability up to 2.7\% that it would hit the Earth on April 13, 2029 \citep{chesley2005potential}. Subsequent observations, however, improved predictions and ruled out the possibility of impact \citep{giorgini2008predicting}. Nevertheless, Apophis will pass within 31,000 km of Earth's surface, closer than geosynchronous satellites \citep{sokolov2012impact}.\\
While it is certain that Apophis will miss the Earth, the exact consequences on the asteroid itself remain uncertain \citep{zhang2020tidal}, with the main question lying in our limited knowledge of Apophis' internal structure. As a matter of fact, the close encounter with the Earth will subject Apophis to significant tidal torques. Earth's gravitational interactions with Apophis will likely modify its physical and dynamical properties, providing valuable insights into asteroid cohesion and strength \citep{souchay2014rotational, valvano2022apophis}. Depending on those, the asteroid could experience different outcomes, from measurable seismic waves and real-time surface disturbance \citep{demartini2019using}, to local surface effects \citep{scheeres2005abrupt}, up to the complete surface reshaping \citep{yu2014numerical,kim2023tidal}. For this reason, this encounter presents an unprecedented planetary defense and science opportunity, and a mission to Apophis has the potential to revolutionize our understanding of the internal structure of asteroids, which is crucial for effective planetary defense missions \citep{binzel2021apophis}.\\
As direct consequence of this celestial event, NASA has recently announced that Apophis will be the selected target of Osiris-REx extended mission \citep{dellagiustina2022osiris}. On the other hand, ESA proposed the mission RAMSES (Rapid Apophis Mission for SEcurity and Safety) having the aim to reach Apophis before the Earth close encounter\footnote{\url{https://esastar-publication-ext.sso.esa.int/ESATenderActions/details/56945} (last accessed: July 7, 2023)}. Its objective is to accompany the asteroid during the fly-by in order to determine the physical and dynamical changes induced by the interaction with the gravity of the Earth. \\
In the context of the RAMSES mission, the present work aims to scrutinize the reachability of the Apophis asteroid. Reachability analysis is an essential task in the preliminary assessment of asteroid missions \citep{wagner2015target,machuca2020high,topputo2021envelop}. In this work, all the possible trajectories departing from the Earth and reaching Apophis before the Earth fly-by are computed and their overall costs are estimated. The best suitable options are identified, and relevant figures that support the system design are evaluated. The variation of the required propellant mass with the departure date and Time of Flight (ToF) have to be produced, with thousands of trajectories to be computed. For this reason, a fast and reliable approach is necessary, in order to have the full envelope of the transfers to Apophis in a relatively small time. To this aim, a two-layer approach has been employed. It first exploits the flexibility of a sequential convex programming algorithm and then an indirect method, the latter guaranteeing the optimality of the solutions.\\
The paper is structured as follows. Section \ref{sec:problems_formulation} presents an overview of the assumptions made on the mission and on the platform. Section \ref{sec:approach} shows the approach used to fast solve the Apophis reachability problem, with details on the employed methodology. The results are presented in Section \ref{sec:results}. Conclusions are drawn in Section \ref{sec:conclusions}.

\section{Mission overview and assumptions}
\label{sec:problems_formulation}

RAMSES is a small satellite mission that aims at the characterization of the dynamical and physical properties of asteroid Apophis before its close encounter with the Earth, on April 13, 2029, and their changes during the fly-by. Considering the Earth fly-by's close temporal proximity, the whole mission design and implementation is strictly driven by time constraints. As a matter of fact, in order to properly map the pre-fly-by characteristics of Apophis, the spacecraft shall reach the asteroid at least 2 months before the encounter. On the other hand, a time frame of at least 3 years is required to design and manufacture the spacecraft, placing the earliest launch date on November 1, 2026. These stringent time constraints open the possibility to use the electric thruster as main propulsion system. Indeed, even though electric propulsion systems require more complex operations, they allow wider departure and arrival windows. In addition, the amount of propellant they require is significantly lower and single-point failure effects are not as catastrophic as they would be when chemical propulsion is considered, since maneuvers are not strongly time-driven. Overall, electric propulsion is a suitable solution for a fast mission to Apophis.\\
The probe wet initial mass $m_0$ has been assumed to be \qty{500}{\kilogram}, with \qty{73}{\kilogram} of available propellant $m_p$. The dry mass represents more than the 85\% of the total mass, corresponding to a typical value for this class of spacecraft. The selected values for the maximum-thrust-to-initial-mass ratio $T_{\text{max}}/m_0$ and for the specific impulse $I_\mathrm{sp}$ for the electric propulsion subsystem are \qty{1.2 e-4}{\meter\per\second\squared} and \qty{1500}{\second}, respectively, which are compatible with state-of-the-art Hall-effect thrusters \citep{dannenmayer2009elementary}. Accordingly, the maximum value for the thrust $T_\mathrm{max}$ is \qty{60}{\milli\newton}.\\
The spacecraft is inserted directly into an interplanetary transfer to Apophis by a launcher. Considering the characteristics of Ariane 6.2 and 6.4 launchers\footnote{Ariane 6 User Manual I2R0: \url{https://www.arianespace.com/wp-content/uploads/2021/03/Mua-6_Issue-2_Revision-0_March-2021.pdf} (last accessed: July 7, 2023)}, the infinity velocity $v_{\infty}$ at the Earth sphere-of-influence interface has been constrained to be lower than \qty{4}{\kilo\meter\per\second}, with free declination $\delta$ and right ascension $\alpha$ in the J2000 reference frame. The analysis considers a two-body problem dynamics, with the Sun as the central body. The main assumptions for the Apophis reachability analysis are listed in Table~\ref{tab:ma_assumptions}. 

\begin{table}[ht!]
    \centering
    \begin{tabular}{lll}
        \toprule
        \toprule
            \textbf{Time Constraints}& Earliest departure date & November 1, 2026\\
            & Max.\ Time of Flight $\text{ToF}_{\text{max}}$ & 800 days\\
            & Latest arrival & February 13, 2029\\
        \midrule
            \textbf{Spacecraft}& $m_0$ & \qty{500}{\kilogram}\\
            & Propulsion & Continuous\\
            & $T_{\text{max}}/m_0$ & \qty{1.2 e-4}{\meter\per\second\squared}\\
            & $I_\mathrm{sp}$ & \qty{1500}{\second}\\
            & $T_\mathrm{max}$ & \qty{60}{\milli\newton}\\
            & $m_p$  & \qty{73}{\kilogram}\\
        \midrule
            \textbf{Launcher}& $v_{\infty}$ & $\leq$ \qty{4}{\kilo\meter\per\second} \\
            & $\alpha$ & $\in [-90, +90]\deg$\\
            & $\delta$ & $\in [-180, +180]\deg$\\
        \bottomrule
        \bottomrule
    \end{tabular}
    \caption{Search space assumptions to perform the reacheability analysis.}
    \label{tab:ma_assumptions}
\end{table}

\section{Approach}
\label{sec:approach}

The procedure followed to analyze the reachability of Apophis has to be fast and reliable, since thousands of fuel-optimal (FO) trajectories must be computed. The output of this analysis is a porkchop plot that shows the propellant mass required to reach Apophis and the reachability of the target for different departure dates and times of flight, considering the assumptions in Table \ref{tab:ma_assumptions}. This assessment has the aim to evaluate the launching conditions and compute the trajectories spanning the whole temporal search space. The approach for this analysis exploits two layers:
\begin{enumerate}[label=\arabic*.]
    \item A direct sequential convex programming (SCP) algorithm \citep{morelli2021robust}, used for each departure date and time of flight. It gives a first assessment of the optimal launching conditions and trajectory, as it allows to easily include the free infinite velocity given by the launcher;
    \item The outputs of the first step are used to feed an indirect method \citep{zhang2015low}, able to guarantee optimality of the solution in terms of propellant mass. 
\end{enumerate}
This two-layer approach is employed to exploit the easy handling of the free launching conditions of the first method and the guaranteed optimality of the second one. The minimum-fuel problem is solved with fixed initial and final boundary conditions, and time of flight. The initial boundary condition is variable in the first step of the methodology and fixed in the second.

\subsection{Convex Optimization with Free Infinite Velocity}
In the first step, the nonconvex low-thrust trajectory optimization problem is solved by considering a sequence of convex subproblems whose solutions eventually converge, under certain hypotheses, to the solution of the original one \citep{malyuta2022convex}. At each iteration, the following problem is solved \citep{Wang2018_fuel,hofmann2022performance}
\begin{subequations}
\begin{align}
    \underset{\vecc{u}(t)}{\text{minimize}} \hphantom{abc} & -w(t_f) + \lambda \max(0, \eta(t)) +  \lambda \|\bm{\nu}(t)\|_1 \label{eq:cost-function} \\
    \text{subject to:} \hphantom{abc} & \dot{\vecc{x}}(t) = \vecc{f}(\vecc{\bar{x}}(t), \vecc{\bar{u}}(t)) + \matt{A}(\vecc{\bar{x}}(t)) \,  (\vecc{x}(t) - \vecc{\bar{x}}(t)) + \matt{B}(\vecc{u}(t)- \vecc{\bar{u}}(t)) + \bm{\nu}(t)  \label{eq:xdot} \\
    & \Gamma(t) \leq T_{\text{max}}  \textrm{e}^{-\bar{w}(t)} \left ( 1 - w(t) + \bar{w}(t) \right )  + \eta(t) \label{eq:constraint-gamma} \\
    & \norm{\bm{\tau}(t)}_2 \leq \Gamma(t) \label{socc} \\
    & \norm{\vecc{x}(t) - \vecc{\bar{x}}(t)}_1 \leq R \label{eq:constraint-tr} \\
    & \vecc{r}(t_i) = \vecc{r}_i, \; \vecc{v}(t_i) = \vecc{v}_i, \; w(t_i) = w_i \label{bcsi}\\
    & \vecc{r}(t_f) = \vecc{r}_f, \; \vecc{v}(t_f) = \vecc{v}_f  \label{bcsf}\\
    & \vecc{x}_l \leq \vecc{x} \leq \vecc{x}_u, \; \vecc{u}_l \leq \vecc{u} \leq \vecc{u}_u \label{bounds}
\end{align}
\label{problem}
\end{subequations}
where $\vecc{x} = [\vecc{r}, \vecc{v}, w]$ and $\vecc{u} = [\tau_x, \tau_y, \tau_z, \Gamma] = [\bm{\tau}, \Gamma]$ are the state and control variables, respectively. The quantities $\eta(t)$ and $\bm{\nu}(t)$ are slack variables to avoid the so-called artificial infeasibility, and the constant parameter $\lambda$ in the objective function is a user-defined weight. The times $t_i$ and $t_f$ are the initial and final transfer times. $R$ is the radius of the trust region assuring that the convexification of the problem is valid. The reader can refer to previous works \citep{bonalli2019gusto} for a more comprehensive explanation of the SCP technique. In Eq.\ \cref{eq:xdot}, the matrices $\vecc{A}$ and $\vecc{B}$ are defined as
\begin{equation}
\label{eq:jacobians}
    \matt{A}(\vecc{\bar{x}}(t)) \defeq \frac{\partial \vecc{f}}{\partial \vecc{x}} \Bigg \rvert_{\vecc{\bar{x}}(t)}, \; \matt{B} \defeq \frac{\partial \vecc{f}}{\partial \vecc{u}} \Bigg \rvert_{\vecc{\bar{u}}(t)} \;
\end{equation}
where
\begin{equation}
    \label{newdynamics}
    \mathbf{f}(\mathbf{x}, \mathbf{u}) = \begin{bmatrix}
                \mathbf{v}(t)\\
                -\frac{\mu}{r^3}\mathbf{r}(t) + \bm{\tau}(t)\\
                -\frac{\Gamma(t)}{I_{\text{sp}}g_0} 
         \end{bmatrix}
\end{equation}
with $r = \norm{\mathbf{r}(t)}_2$. The problem in Eqs.\ \cref{problem} is a Second-Order Cone Program (SOCP) and can be handled by efficient convex solvers. In this work, it has been solved using an Hermite--Simpson discretization scheme \citep{morelli2021robust} and the Embedded COnic Solver (ECOS) \citep{ecos}. \\
In order to account for the additional degree of freedom provided by the launcher, the problem is enhanced as follows:
\begin{enumerate}[label=\arabic*.]
    \item Three additional variables $v_{\infty}^x, v_{\infty}^y$, and $v_{\infty}^z$ are introduced and collected in the vector $\mathbf{v}_{\infty} \in \mathbb{R}^{3}$.
    \item The initial boundary conditions on the velocity of the spacecraft in Eq.\ \cref{bcsi} are expressed as
    \begin{equation}
        \mathbf{v}_i = \mathbf{v}_{\text{E}}(t_i) + \mathbf{v}_{\infty}
        \label{eq:v_inf_initial}
     \end{equation}
     where $\mathbf{v}_{\text{E}}(t_i)$ indicates the velocity of the Earth at the initial time $t_i$.
     \item In J2000, it is possible to express the constraints on the maximum magnitude of the infinite velocity and on its maximum declination in Table \ref{tab:ma_assumptions} as
     \begin{subequations}
         \begin{align}
             & \| \mathbf{v}_{\infty} \|_2 \leq v_{\infty}^{\text{max}} \label{soccla}\\
             & -v_{\infty}^{z,\text{max}} \leq v_{\infty}^z \leq v_{\infty}^{z,\text{max}} \\
             & v_{\infty}^{z,\text{max}} = v_{\infty}^{\text{max}} \sin{\tilde{\delta}}
         \end{align}
         \label{constrlauncher}
     \end{subequations}
 where $\tilde{\delta}$ is the maximum allowable declination. Note that the constraint in Eq.\ \cref{soccla} is a Second-Order Cone Constraint (SOCC). Therefore, considering the constraints in Eqs.\ \cref{constrlauncher} still makes the problem in Eqs.~\cref{problem} convex. It is worth to explicitly mention that the right ascension of the infinite velocity is instead free to vary. 
 \end{enumerate}
 Although the SCP algorithm is rather robust \citep{hofmann2022performance}, the solution of a given problem depends on the initial guess. Therefore, to obtain an homogeneous porkchop plot, a continuation method has been used, which considers the solutions of already-solved problems as initial guess for their neighbours. This also helps reducing significantly the computational time \citep{topputo2021envelop}. The continuation procedure can be summarized as follows.
\begin{enumerate}[label=\arabic*.]
\item First, the trajectory associated with the earliest departure date and longest duration is computed with a simple shape-based initial guess. This trajectory corresponds to the green box of the matrix in Fig.~\ref{fig:porkscheme}. 
\item The search space is discretized using time grids. In particular, a number $M$ of times of flight and a number $N$ of departure dates are considered.
\item Then, for all $i = M-1, \dots, 1$ and earliest departure date, the solution of the (already solved) problem associated with higher time of flight is interpolated to obtain the initial guess. Since an Hermite--Simpson discretization scheme is used, the state in each trajectory segment is a third-order polynomial. If the step in the time of flight direction is fairly small, the state of Apophis correspondent to the cases $i-1$ and $i$ is similar, and therefore the considered trajectory represents a consistently good initial guess. In the case when the optimal control problem $(i-1,\ 1)$ did not converge, a cubic-based initial guess is instead used for the problem $(i,\ 1)$. 
\item Once the trajectories associated with all the time of flights and first departure date have been computed, an additional continuation is performed. In particular, for all departure dates $j = 2, \dots, N$, the solution of the problem with time of flight $i$ and departure date $j-1$ is used from the new departure date $j$ to the arrival date associated with the trajectory $(i,\ j-1)$ to obtain the initial guess for the problem $(i,\ j)$. In case the case $(i,\ j-1)$ did not converge, a shape-based initial guess is used instead. 
\item For each of the departure dates, the FO trajectory with lower time of flight that reached convergence is finally used as initial guess to solve the time-optimal (TO) problem as well \citep{morelli2021robust}. This continuation corresponds to the last separated row of the matrix in Fig.~\ref{fig:porkscheme}.
\end{enumerate}
In order to further speed-up the convergence process, the number of nodes used by the algorithm to solve the problems is considered to be directly proportional to the time of flight. In particular, a number of nodes $P_{\textrm{max}} = 150$ for the largest time of flight $\text{ToF}_{\textrm{max}} = 800$ days is chosen, and therefore
\begin{equation}
    P = \left\lceil P_{\textrm{max}} \frac{\text{ToF}}{\text{ToF}_{\textrm{max}}} \right\rceil
\end{equation}
\begin{figure}[h!]
	\centering
    \includegraphics[width=0.5\textwidth]{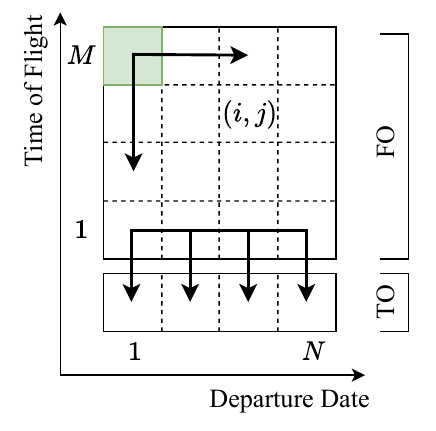}
    \caption{Continuation scheme for the feasibility plot adopted in the two layers.}
    \label{fig:porkscheme}
\end{figure}
\noindent To properly detect the variations in propellant mass $m_p$ and having a more efficient continuation, a variable step grid in the time-of-flight dimension is exploited. From ToF of 244 days up to 460 days, the time step has been considered to be 1 day, while a 20-day interval is considered from 460 on. This step is necessary to refine the porkchop in the zone closer to the TO solution, where both the convex and the indirect solvers benefit from denser discretization grids. Contrarily, the time step for the departure dates is fixed and set to 5 days. This grid requires about 16,000 trajectories to be computed, confirming that a fast methodology is needed to assess the feasibility of a low-thrust mission to Apophis.

\subsection{Indirect Optimization Refinement}
\label{sub_sec:indirect_step}

The results obtained with the convex step are used as inputs for the indirect formulation and the generation of the final porkchop. Differently from direct methods, indirect ones aim at satisfying the necessary optimality conditions of the low-thrust trajectory optimization problem, which are derived through the calculus of variations \citep{kechichian1997optimal}. The methodology developed in previous works \citep{zhang2015low} is used. It solves the shooting problem that stems from the imposition of the necessary conditions as specified here after. In particular, the augmented dynamical equations of states $[\mathbf{r}, \mathbf{v}, m]$ and costates $[\boldsymbol{\lambda}_r, \boldsymbol{\lambda}_v, \lambda_m]$ for the two-body problem are considered \citep{wang2022indirect}
\begin{equation}
    \dot{\mathbf{y}} = \mathbf{F}(\mathbf{y}) \Rightarrow \begin{cases}
    \dot{\mathbf{r}} = \mathbf{v} \\ 
    \dot{\mathbf{v}} = -\frac{\mu}{r^3}\mathbf{r} + u\frac{T_{max}}{m}\boldsymbol{\alpha} \\ 
    \dot{m} = -u\frac{T_{\text{max}}}{I_\mathrm{sp} g_0} \\
    \dot{\boldsymbol{\lambda}}_r = -\frac{3\mu}{r^5}(\mathbf{r} \cdot \boldsymbol{\lambda}_v)\mathbf{r} + \frac{\mu}{r^3}\boldsymbol{\lambda}_v\\
    \dot{\boldsymbol{\lambda}}_v = -\boldsymbol{\lambda}_r\\
    \dot{\lambda}_m = -u \frac{ T_\mathrm{max}}{m^2}\lambda_v
    \end{cases}
    \label{eq:aug_sys_eq_dynamics}
\end{equation}
where $u(t) \in [0, 1]$ is the throttle factor and $\boldsymbol{\alpha}(t)$ is the thrust direction vector, and have to be imposed to their optimal value $\mathbf{u}^*(t) = [u^*(t), \boldsymbol{\alpha}^*(t)]$ such that the Pontryagin Maximum Principle \citep{bryson1969applied} is respected. To integrate these equations, the algorithm enforces the initial conditions at $t_i$ for both the TO and FO problems
\begin{equation}
    \mathbf{r}(t_i) - \mathbf{r}_{i} = \mathbf{0}, \quad \mathbf{v}(t_i) - \mathbf{v}_{i} = \mathbf{0}, \quad m(t_i) - m_0 = 0
    \label{eq:ini_cond}
\end{equation}
The initial mass $m_0$ is constant and considered equal to 500 kg as per the assumptions in Table~\ref{tab:ma_assumptions}, whereas $\mathbf{r}_i$ is the initial position of the Earth, which varies with the departure date considered in each point of the search space mesh. The initial velocity is defined from the results of the convex optimization step. In particular, according to Eq.~\eqref{eq:v_inf_initial}, it is
\begin{equation}
\mathbf{v}_i = \mathbf{v}_{\text{E}}(t_i) + \mathbf{v}_{\infty} = 
\begin{bmatrix}
    v_{\text{E},x}(t_i) + v_{\infty} \cos{{\alpha}}\cos{\delta}\\ 
    v_{\text{E},y}(t_i) + v_{\infty} \sin{{\alpha}}\cos{\delta}\\
    v_{\text{E},z}(t_i) + v_{\infty} \sin{\delta}
    \end{bmatrix}
    \label{eq:initial_vel_constr_lt20}
\end{equation}
where $v_{\infty}$, $\alpha$, and $\delta$, come from the results of the first convex layer.\\
To obtain the solutions in the search space mesh, the same continuation scheme described in Fig.~\ref{fig:porkscheme} is exploited for the indirect formulation. We first solve the FO problem: $\mathbf{y}(t) = \boldsymbol{\varphi}(\mathbf{y}_i, t_i, t)$ being the solution flow integrated from $t_i$ to a generic time instant $t$, the FO shooting problems aim to find $\boldsymbol{\lambda}_i^{*} = \boldsymbol{\lambda}^{*}(t_i)$ such that the solution at the final time $t_f$, namely $\mathbf{y}(t_f) = \boldsymbol{\varphi}([\mathbf{x}_i, \, \boldsymbol{\lambda}_i^*], t_i, t_f)$, satisfies the boundary conditions
\begin{equation} 
\begin{bmatrix}
    \mathbf{r}(t_f) - \mathbf{r}_f \\ 
    \mathbf{v}(t_f) - \mathbf{v}_f\\
    \lambda_m(t_f) 
    \end{bmatrix} = \mathbf{0}
    \label{eq:boundary_conds}
\end{equation}
where $\mathbf{r}_f$ and $\mathbf{v}_f$ are the known final position and velocity of Apophis, respectively, which depend on on the arrival date considered for the specific point of the search space mesh. In the integration of the flow, it can be proved \citep{zhang2015low} that the optimal thrust vector can be expressed as $\boldsymbol{\alpha}^* = -\frac{\boldsymbol{\lambda_v}}{\lambda_v}$, whereas the optimal throttle factor $u^*$ has the following structure: 
\begin{equation}
    u^* = \begin{cases}
        0 \quad &\mbox{if} \quad S_{FO} > 0 \\
        \in [0, 1] \quad &\mbox{if} \quad S_{FO} = 0 \\
        1 \quad &\mbox{if} \quad S_{FO} < 0 
    \end{cases}
    \label{eq:optimal_thrust_FO}
\end{equation}
where $S_{FO}$ is the FO switching function, $S_{FO} = 1 - \lambda_m  -\lambda_v \frac{I_{sp}g_0}{m}$. When solving the TO problem, the objective is to find the pair $[\lambda_i^{*}, \, t_f^*]$ such that $\mathbf{y}(t_f^*) = \boldsymbol{\varphi}([\mathbf{x}_i, \, \boldsymbol{\lambda}_i^*], t_i, t_f^*)$ obtained by the integration of Eq.~\eqref{eq:aug_sys_eq_dynamics} satisfies the boundary conditions
\begin{equation} 
\begin{bmatrix}
    \mathbf{r}(t_f) - \mathbf{r}_{T}(t_f) \\ 
    \mathbf{v}(t_f) - \mathbf{v}_{T}(t_f)\\
    \lambda_m(t_f) \\
    H_t(t_f) - \boldsymbol{\lambda}_r(t_f) \cdot\mathbf{v}_f - \boldsymbol{\lambda}_v(t_f) \cdot\mathbf{a}_f
    \end{bmatrix} = \mathbf{0}
    \label{eq:boundary_conds_TO}
\end{equation}
where $H_t$ is the Hamiltonian of the TO problem, i.e., $H_t = 1 + \boldsymbol{\lambda}^T \cdot\mathbf{f}$ with $\mathbf{f} = \mathbf{f}(\mathbf{x}(t), \mathbf{u}(t))$ being the dynamics of the spacecraft, and $\mathbf{a}_f = \dot{\mathbf{v}}_f$ is the acceleration of Apophis at the final time. In the integration for the TO problem, the optimal $\boldsymbol{\alpha}^*$ and $u^*$ follow the same logic of the FO problem, apart from the switching function which is replaced with the TO switching function $S_{TO} = - \lambda_m  -\lambda_v \frac{I_{sp}g_0}{m}$. To solve the TO problem the values of $v_{\infty}$, $\alpha$, and $\delta$ obtained from the convex layer are exploited, as well as the value of the ToF, which is used as initial guess for the optimal $t_f^*$. %

\section{Results}
\label{sec:results}
Considering the  mission constraints in Table~\ref{tab:ma_assumptions} and the approach  in Section~\ref{sec:approach}, three different analyses are carried out. First, a reachability assessment over the full departure and arrival windows is performed with a nominal thrust-to-mass ratio. Three convenient launch windows are identified and therefore three correspondent representative trajectories are analysed. Finally, it is shown that a fourth launch window can open up if engines with higher thrust-to-mass ratios are considered. Details about these analyses are provided in the remainder.

\subsection{Reachability assessment}
\begin{figure}[!htbp]
	\centering
	{\includegraphics[width=0.78\textwidth]{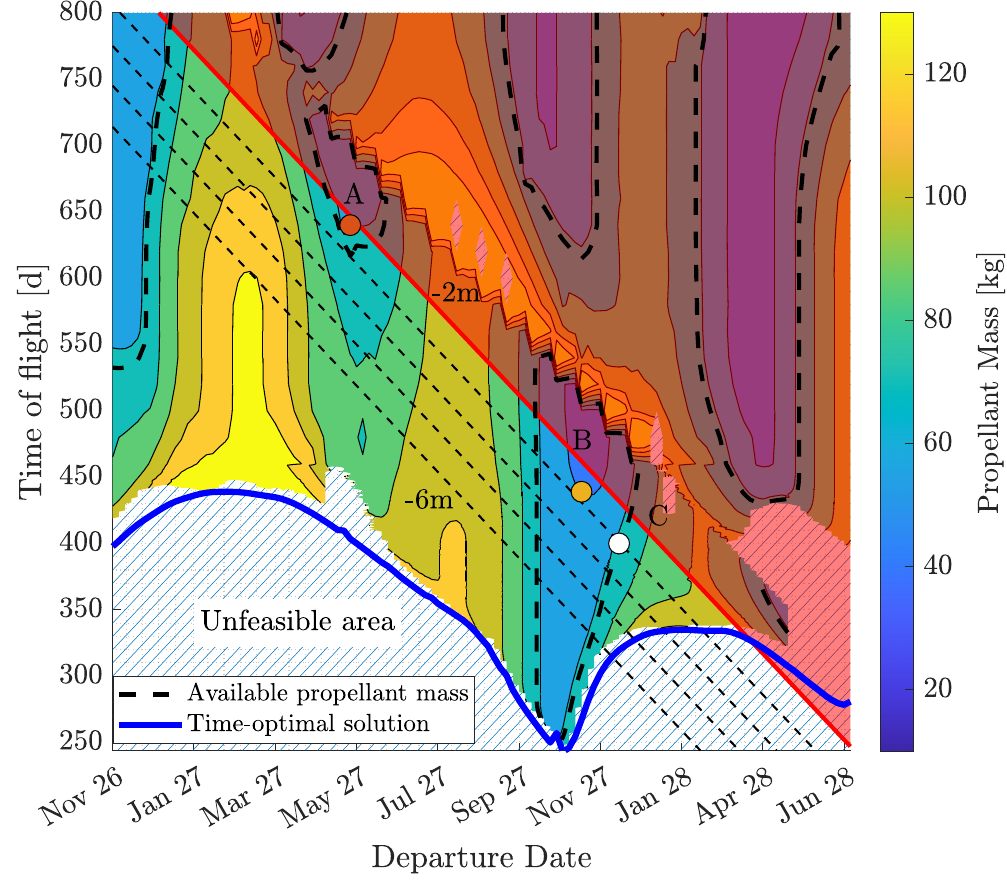}
	\caption{Propellant mass porkchop plot.}	\label{fig:porkchop}}

    \vspace*{\floatsep}
    \centering
	\includegraphics[width=0.78\textwidth]{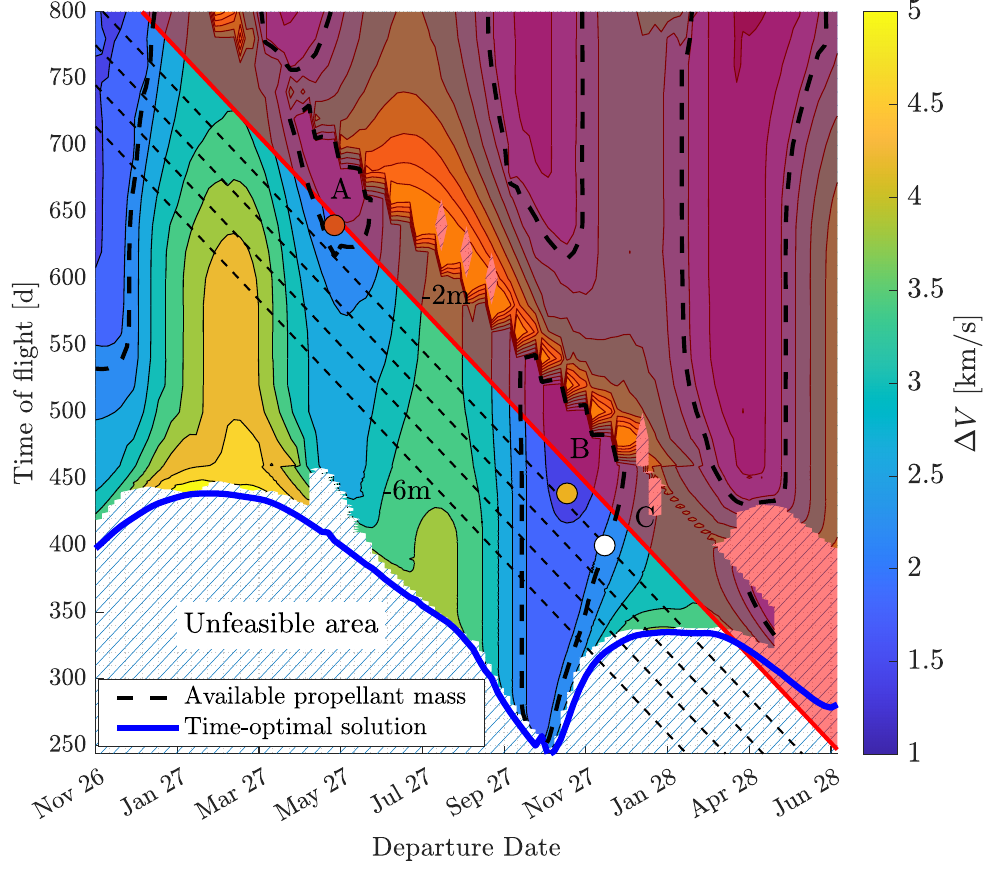}
	\caption{$\Delta V$ porkchop plot.}	\label{fig:porkchop_deltaV}
\end{figure}

\begin{figure}[!ht]
\centering
\subfigure[][{Excess velocity magnitude $v_{\infty}$.}]
{\includegraphics[width=0.48 \textwidth]{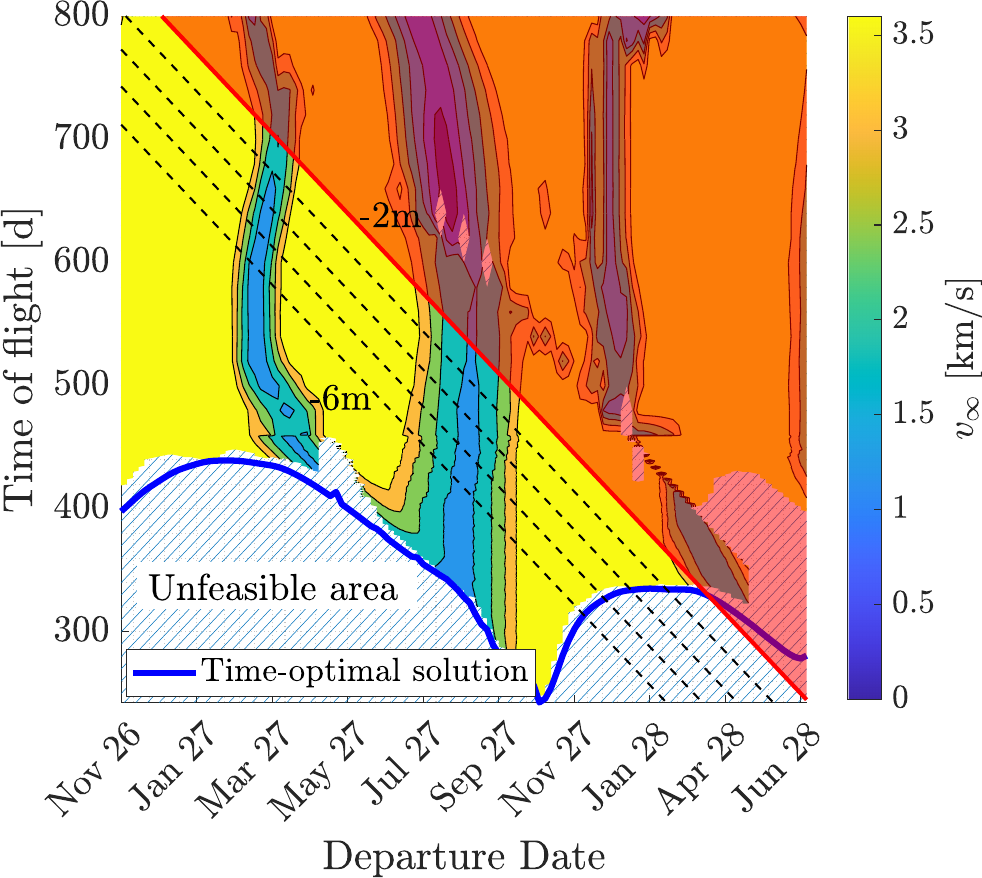}}\\
\subfigure[][{Excess velocity declination $\delta$.}]
{\includegraphics[width=0.48 \textwidth]{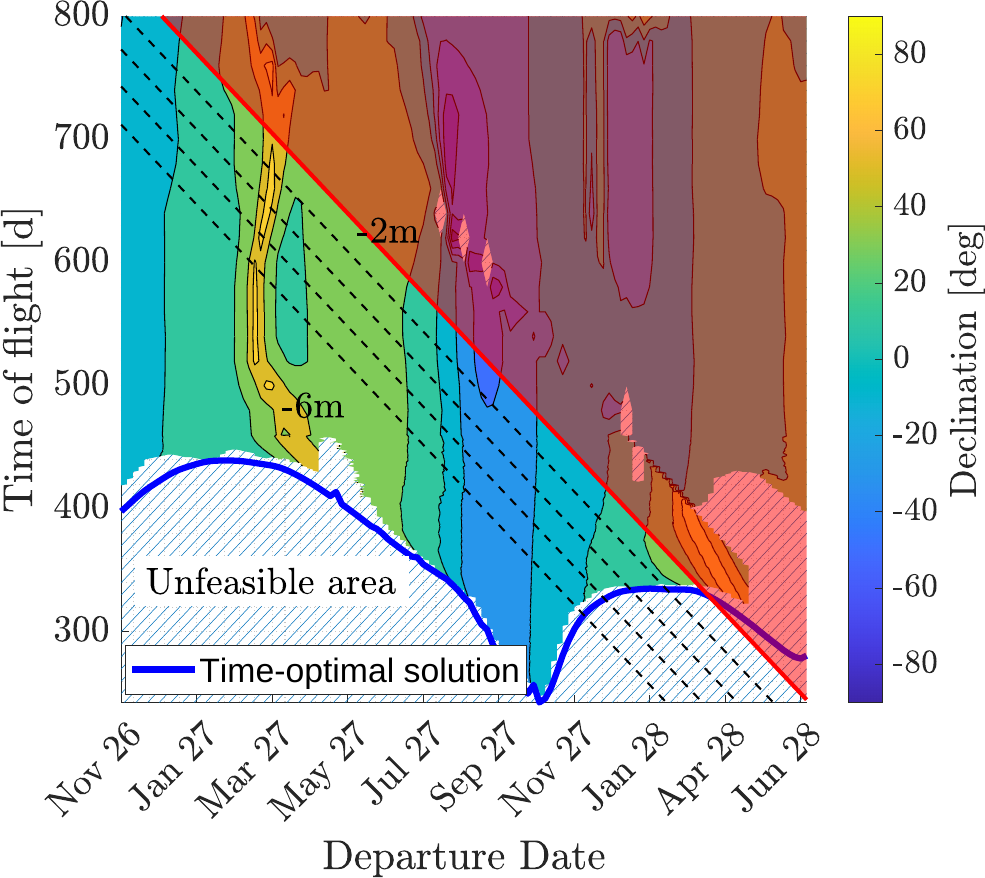}}
\subfigure[][{Excess velocity right ascension $\alpha$.}]
{\includegraphics[width=0.48 \textwidth]{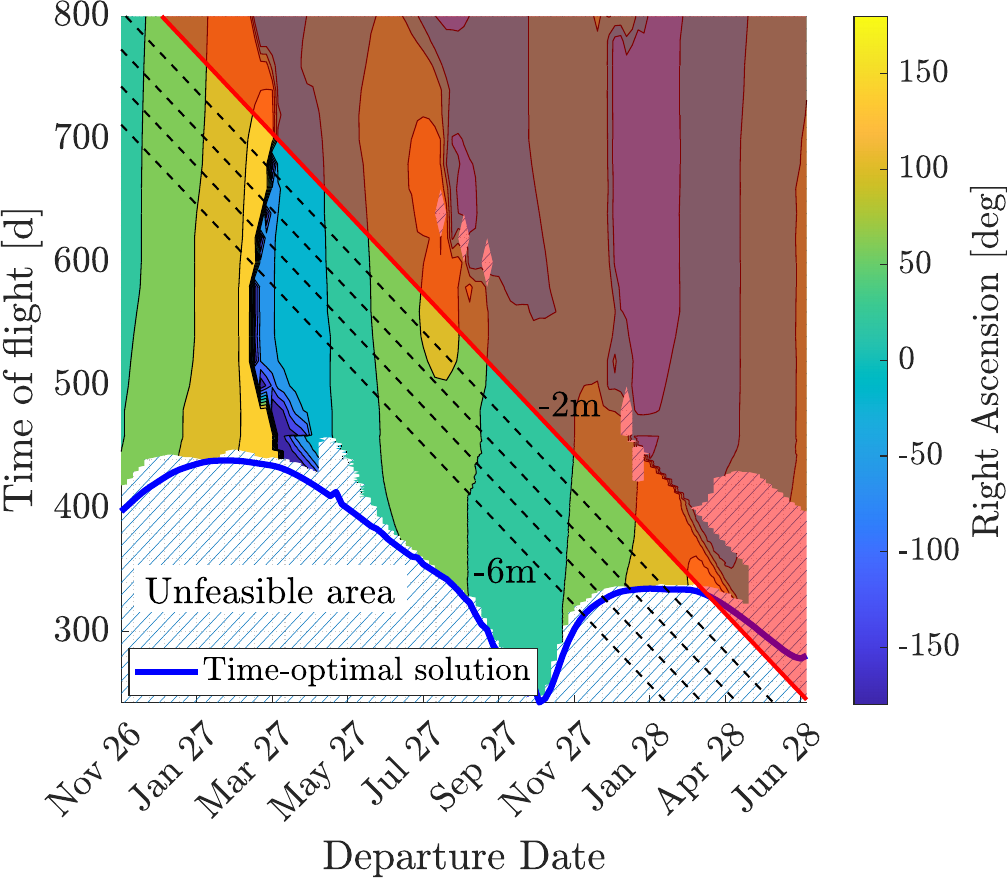}}
\caption{Characteristics of the optimal excess velocity provided by the launcher for each departure date and time of flight.}
\label{fig:Vinf_char}
\end{figure}

Figure~\ref{fig:porkchop} shows the porkchop plot related to the reachability of the Apophis asteroid under nominal conditions. This and the following plots are to be read as follows:
\begin{itemize}
    \item The $x$-axis represents the departure date;
    \item The $y$-axis represents the time of flight;
    \item The color code represents the quantity of interest, indicated on the colorbar on the right of the plot;
    \item The patched red area indicates rendezvous happening after the latest arrival accordingly to Table~\ref{tab:ma_assumptions}, and that are therefore unfeasible;
    \item The oblique dashed segments represent rendezvous dates coinciding with, from up to down respectively, -3, -4, -5, and -6 months from Apophis' closest encounter with the Earth;
    \item The bold blue line indicates time-optimal solutions, i.e., the interplanetary transfers for which the thruster is kept on for the whole time of flight. Note that for a given departure date, solutions with lower times of flight are unfeasible (corresponding to the blue-dashed area);
    \item The dashed black line represents the available propellant of 73 kg.
\end{itemize}
The irregular behaviour at the center of the porkchop is related to the fly-by at the Earth. In that point, a quasi-impulsive velocity change is experienced by the asteroid in the heliocentric reference frame, making it difficult for the optimizer finding a solution for the rendezvous. Figure~\ref{fig:porkchop_deltaV} shows the required $\Delta V$, computed using the Tsiolkovsky equation as $\Delta V = I_{sp}g_0 \ln{\frac{m_0}{m_f}}$.\\
Results show that, under the considered hypotheses, there are three feasible launch windows, namely:
\begin{itemize}
    \item A first one, with launch dates between November and December 2026 and times of flight that span from 540 and 800 days approximately (upper-left part of Fig.~\ref{fig:porkchop});
    \item A second one, with launch dates between April and May 2027 and times of flight of 620-660 days;
    \item A third one, consistently larger, with launch dates from September to November 2027, and times of flight spanning from 250 to 500 days.
\end{itemize}
Figures~\ref{fig:Vinf_char} present the optimal magnitude, declination, and right ascension, respectively, of the excess velocity of the spacecraft with respect to the Earth in the equatorial, heliocentric reference frame (J2000) as a function of the departure date and time of flight. In all the aforementioned windows, the norm of the excess velocity reaches the maximum allowable value. Moreover, the optimal declination of the excess velocity in the regions of interest never assumes large values and it is included between \qty{+-40}{\deg}, values which are compatible with the Ariane 6.2 and 6.4 launchers.

\begin{figure}[!ht]
\centering
\subfigure[][{CPU time for the convex layer.}]
{\includegraphics[width=0.48 \textwidth]{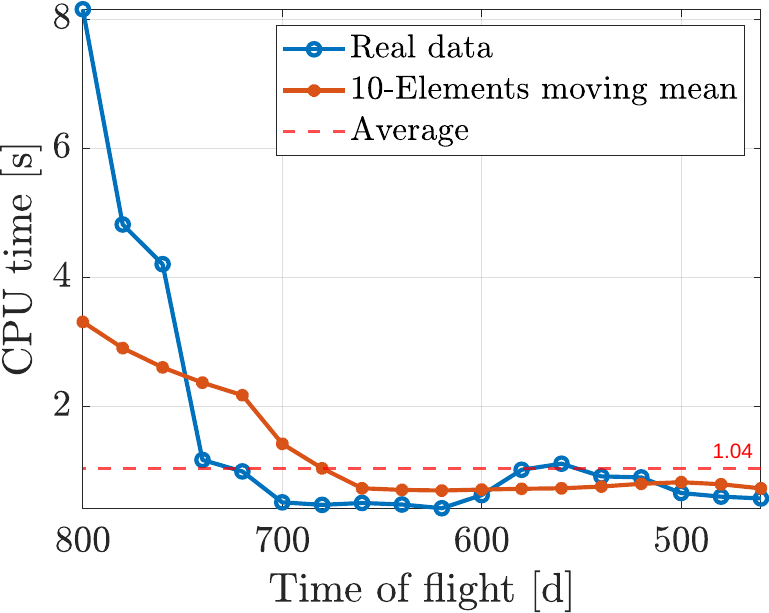}}
\subfigure[][{CPU time for the indirect layer.}]
{\includegraphics[width=0.48 \textwidth]{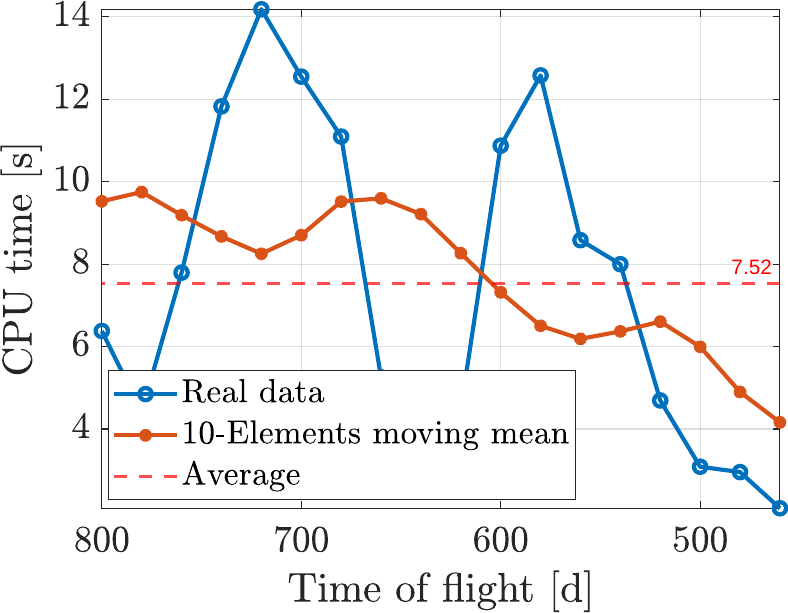}}
\caption{CPU time required for all converged times of flight at the earliest departure date.}
\label{fig:CPUtimes}
\end{figure}

\noindent Figures~\ref{fig:CPUtimes} show the computational time required for the two layers of the methodology. Results refer to the first departure date. A single optimal solution requires at most \qty{15}{\second} to run both the layers, with a mean runtime of less than \qty{9}{\second}. Additionally, it is worth noting that the continuation scheme proves beneficial in expediting the identification of solutions with lower ToF. Typically, these solutions require longer computation times due to their proximity to the time-optimal solution. As a matter of fact, the convex layer requires less than \qty{1}{\second} to compute the optimal solution. The indirect layer requires slightly longer times due to the internal continuation from the energy- to the fuel-optimal problem \citep{zhang2015low}. In conclusion, only few hours are required to compute the porkchop plot and complete the reachability analysis.

\subsection{Candidate Trajectories}

\begin{figure}[!ht]
    \centering
    \subfigure[][{Thrust profile of solution A.}]
    {\includegraphics[width=0.32 \textwidth]{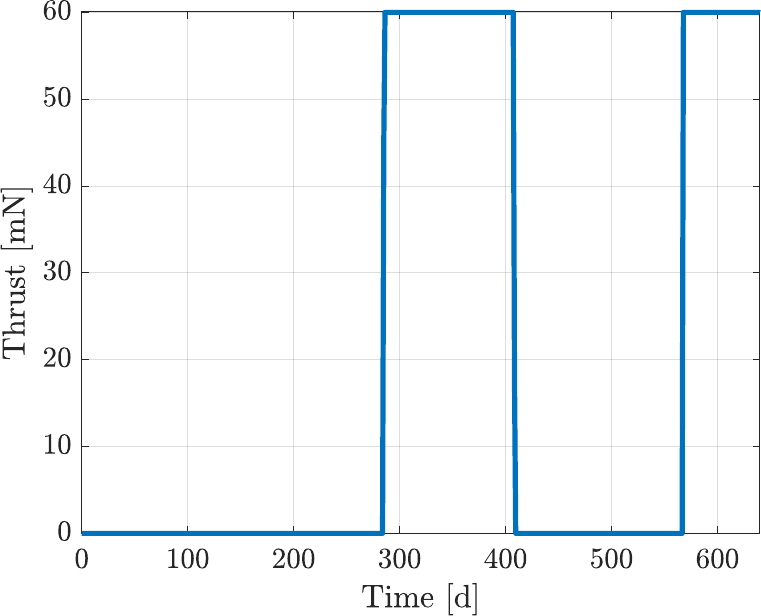}}
    \subfigure[][{Thrust profile of solution B.}]
    {\includegraphics[width=0.32 \textwidth]{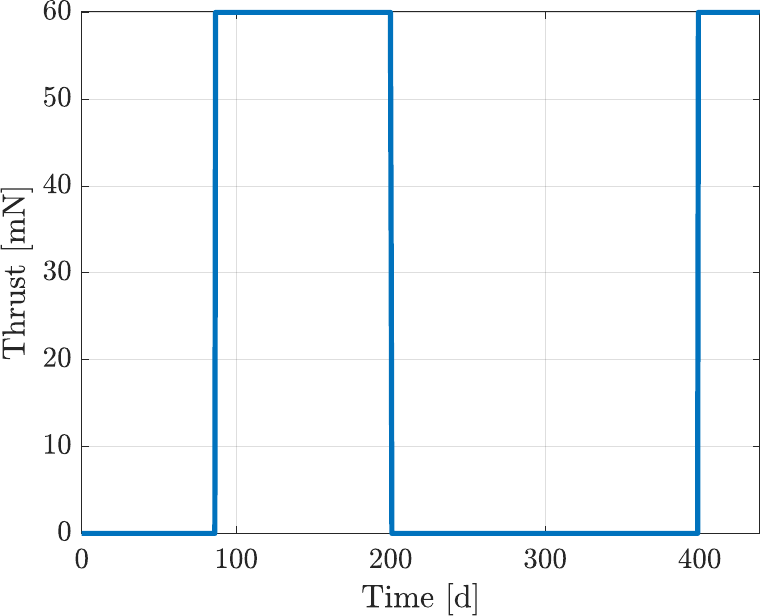}}
    \subfigure[][{Thrust profile of solution C.}]
    {\includegraphics[width=0.33 \textwidth]{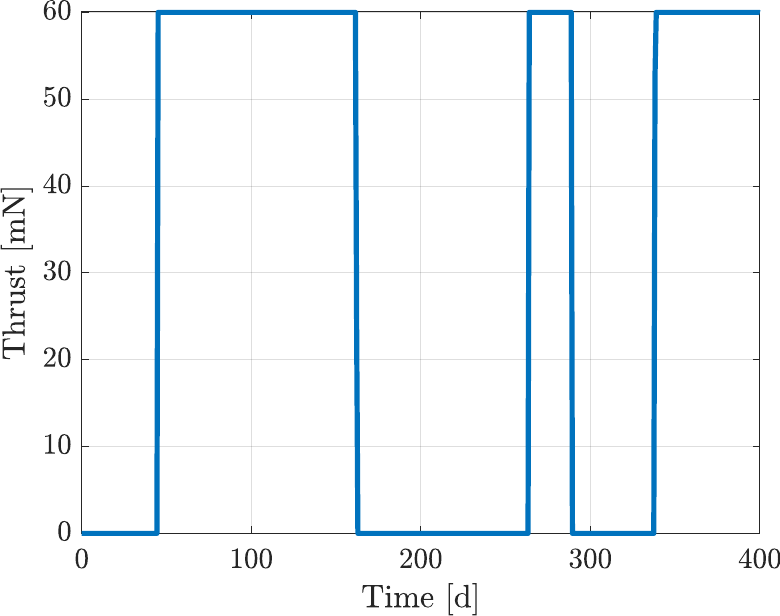}}

    \caption{Thrust profiles of trajectories A, B, and C.}
    \label{fig:selected_trajectories_profiles}
\end{figure}

\begin{figure}[!ht]
\centering
\subfigure[][{Spacecraft-Apophis distance.}]
{\includegraphics[width=0.48 \textwidth]{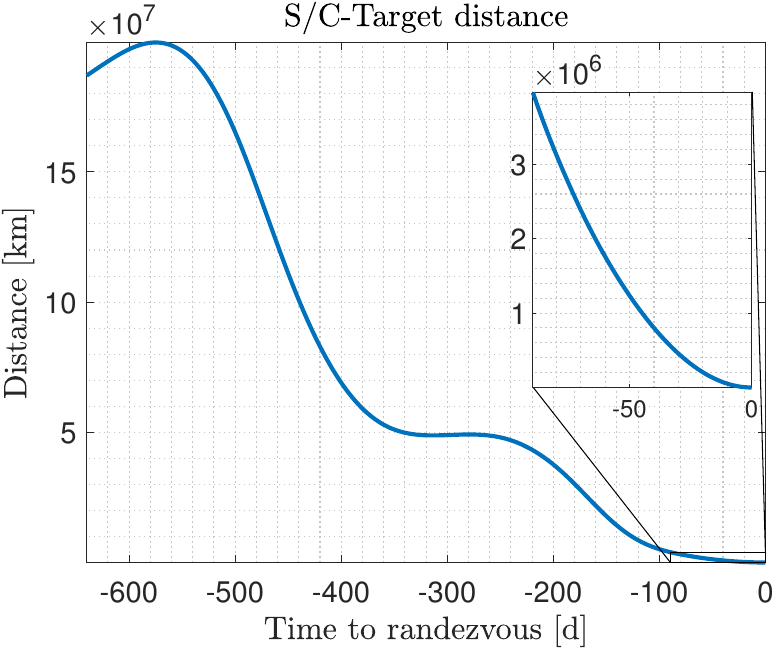}}
\subfigure[][{Spacecraft-Earth distance.}]
{\includegraphics[width=0.48 \textwidth]{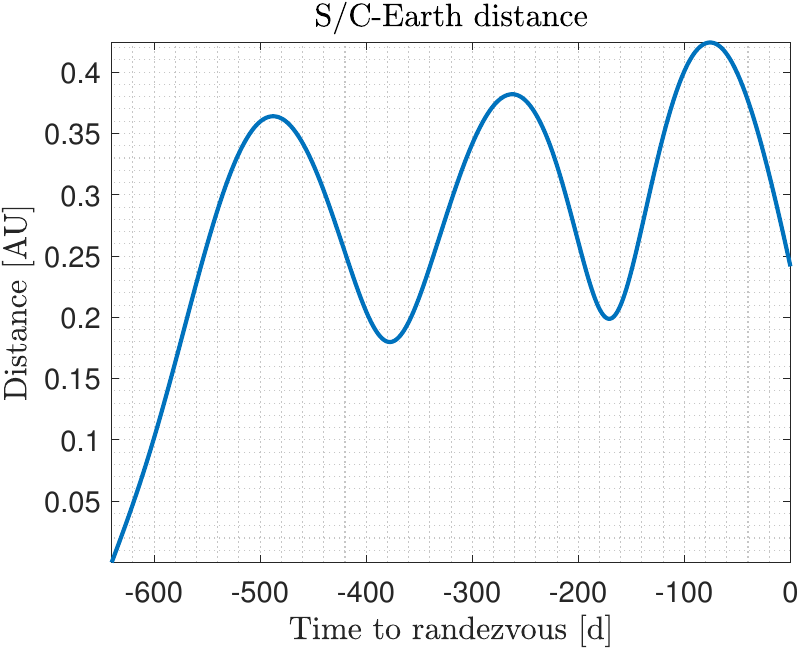}}\\
\subfigure[][{Spacecraft-Sun distance.}]
{\includegraphics[width=0.48 \textwidth]{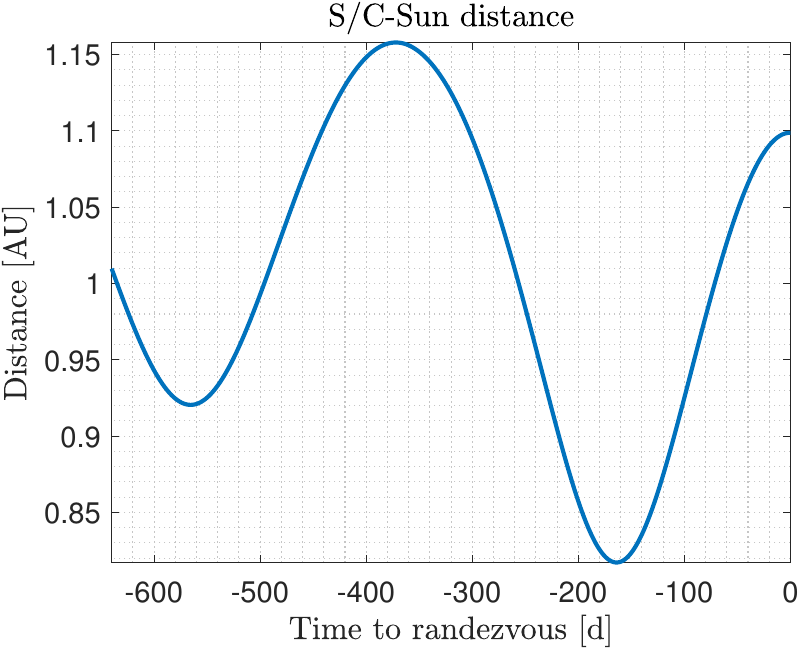}}
\subfigure[][{Target phase angle.}]
{\includegraphics[width=0.48 \textwidth]{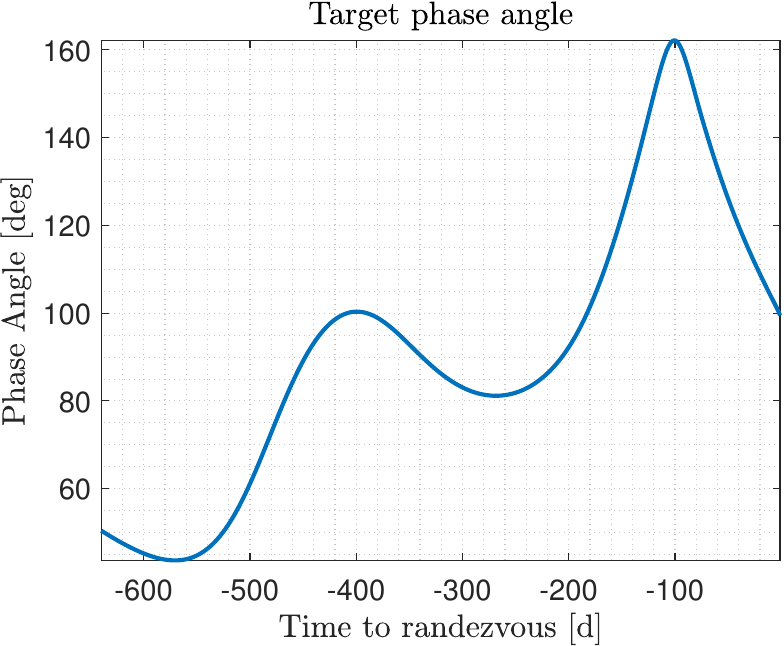}}\\
\subfigure[][{Trajectory representation.}]
{\includegraphics[width=0.7 \textwidth]{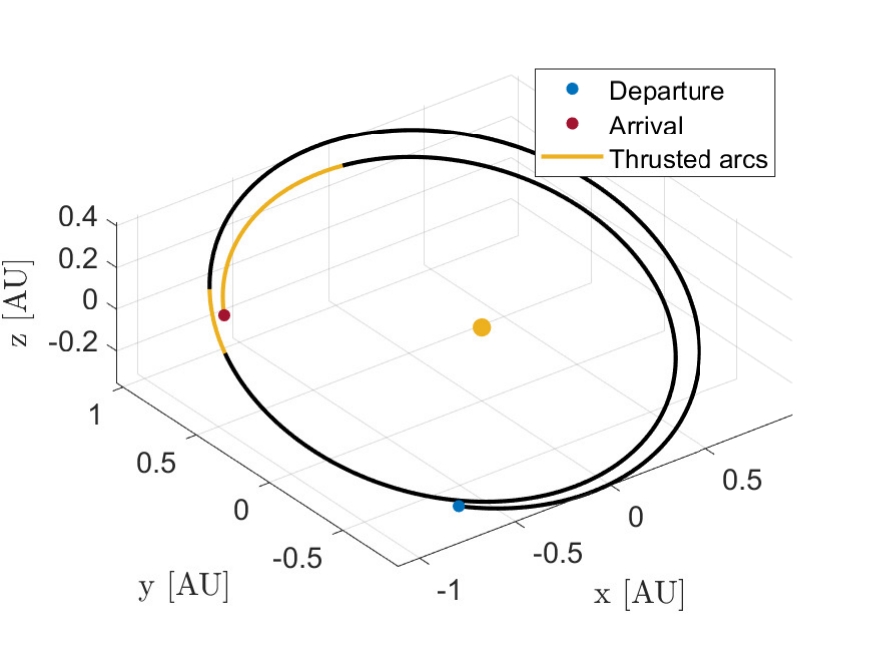}}
\caption{Geometrical analysis and representation in the J2000 reference frame of the solution A.}
\label{fig:geom_may_traj}
\end{figure}

\begin{figure}[!ht]
\centering
\subfigure[][{Spacecraft-Apophis distance.}]
{\includegraphics[width=0.48 \textwidth]{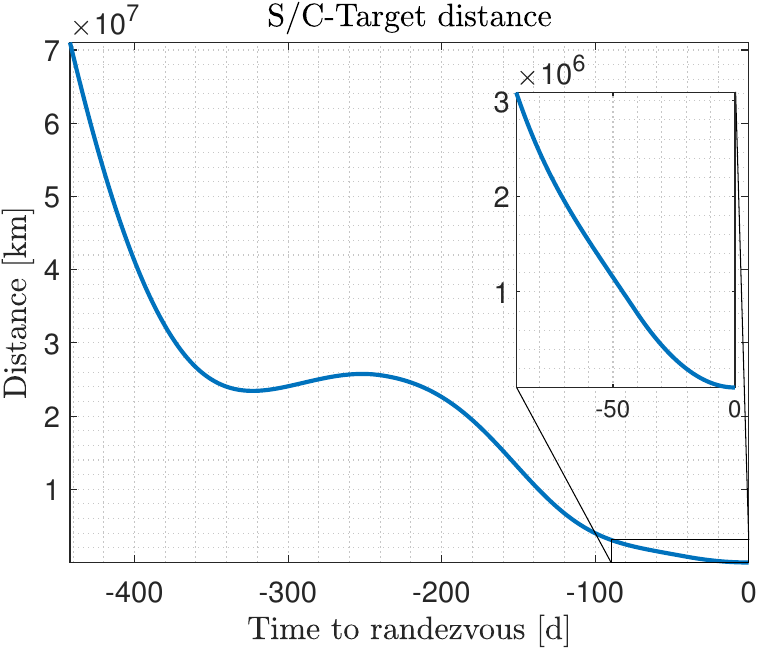}}
\subfigure[][{Spacecraft-Earth distance.}]
{\includegraphics[width=0.48 \textwidth]{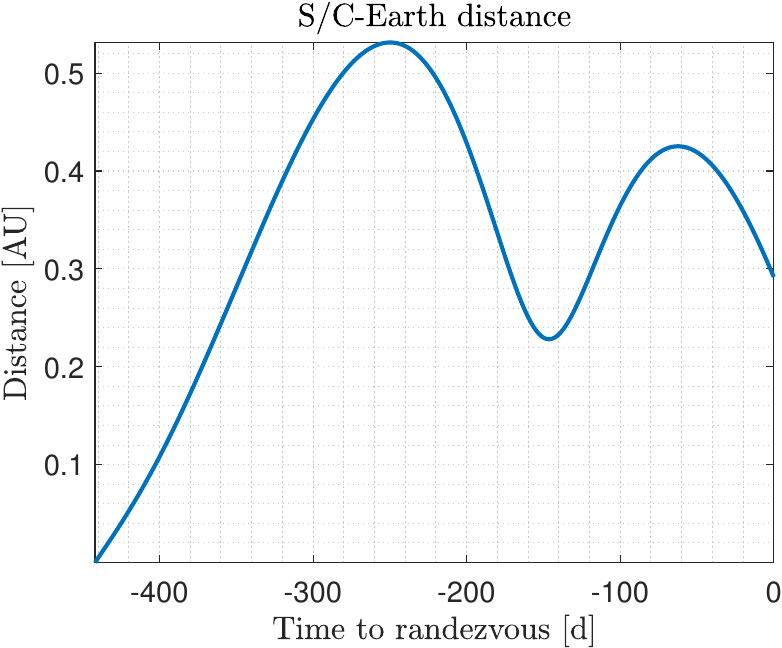}}\\
\subfigure[][{Spacecraft-Sun distance.}]
{\includegraphics[width=0.48 \textwidth]{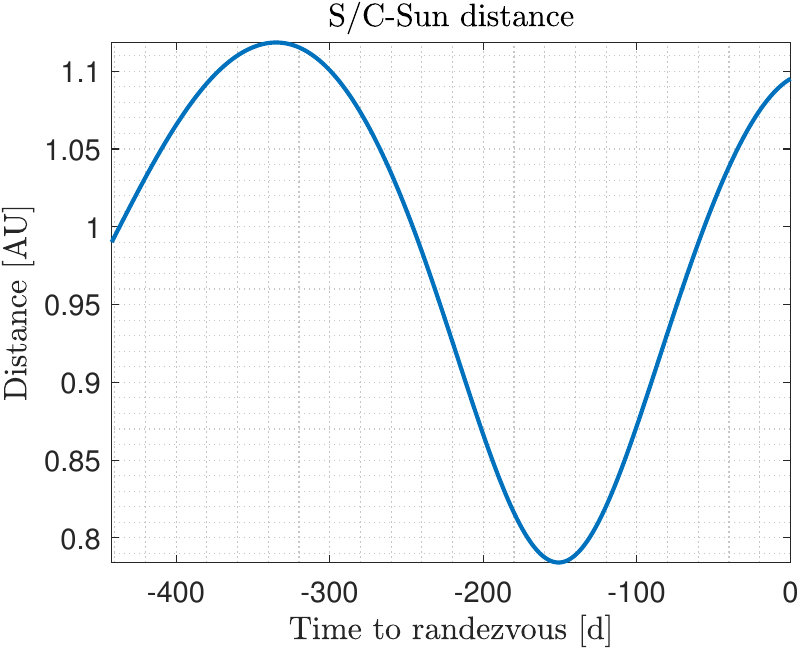}}
\subfigure[][{Target phase angle.}]
{\includegraphics[width=0.48 \textwidth]{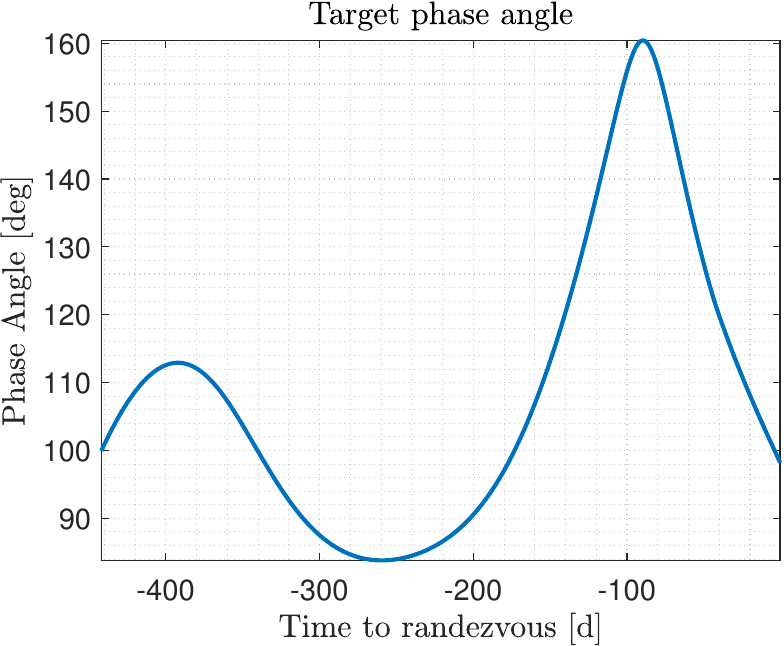}}\\
\subfigure[][{Trajectory representation.}]
{\includegraphics[width=0.7 \textwidth]{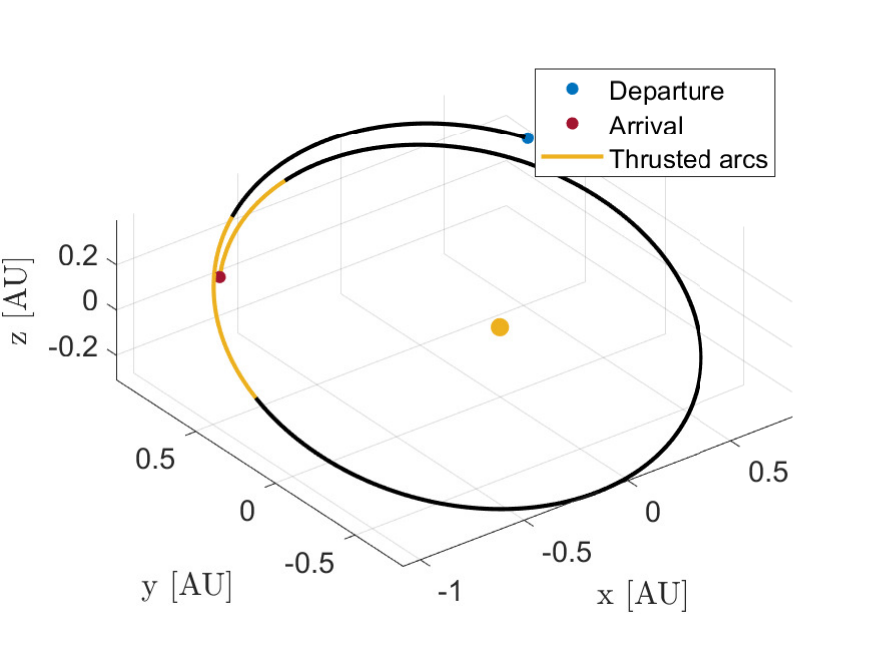}}
\caption{Geometrical analysis and representation in the J2000 reference frame of the solution B.}
\label{fig:geom_nov_traj}
\end{figure}

\begin{figure}[!ht]
\centering
\subfigure[][{Spacecraft-Apophis distance.}]
{\includegraphics[width=0.48 \textwidth]{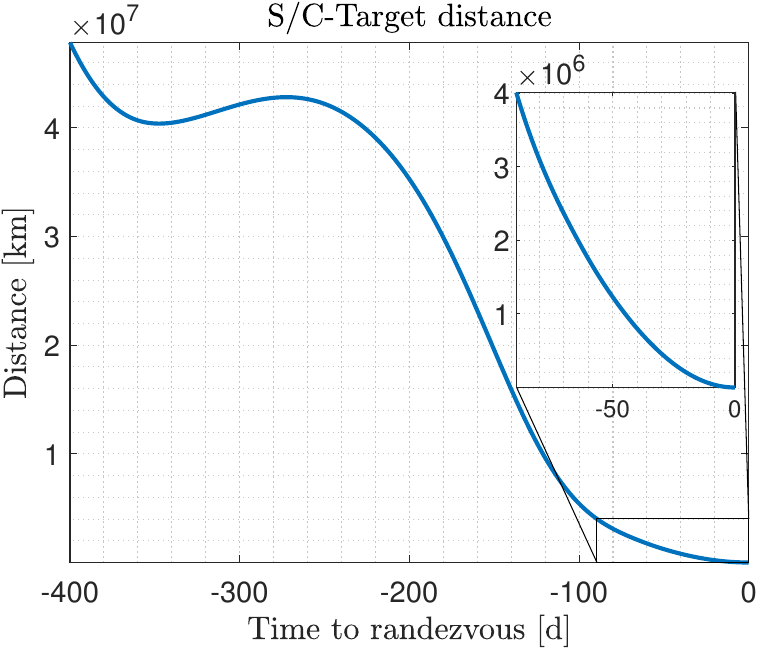}}
\subfigure[][{Spacecraft-Earth distance.}]
{\includegraphics[width=0.48 \textwidth]{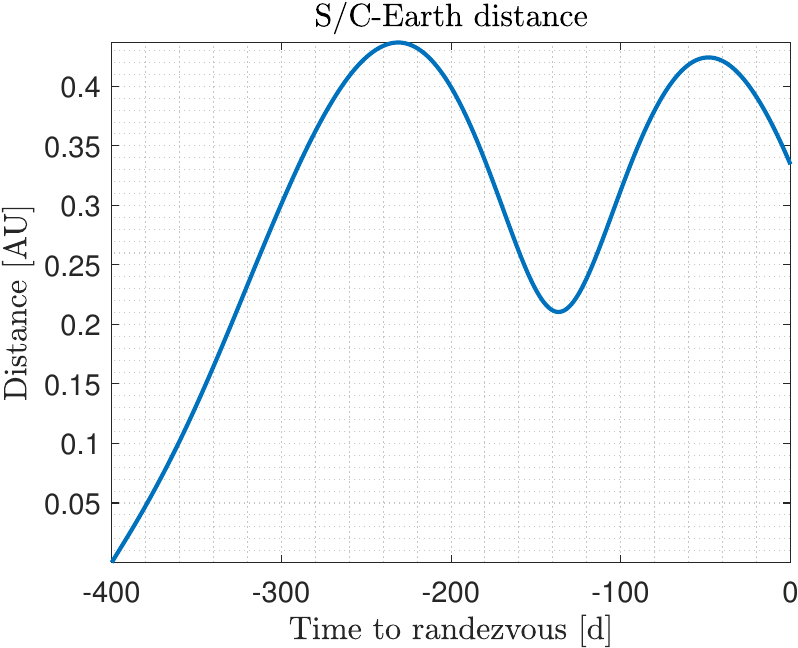}}\\
\subfigure[][{Spacecraft-Sun distance.}]
{\includegraphics[width=0.48 \textwidth]{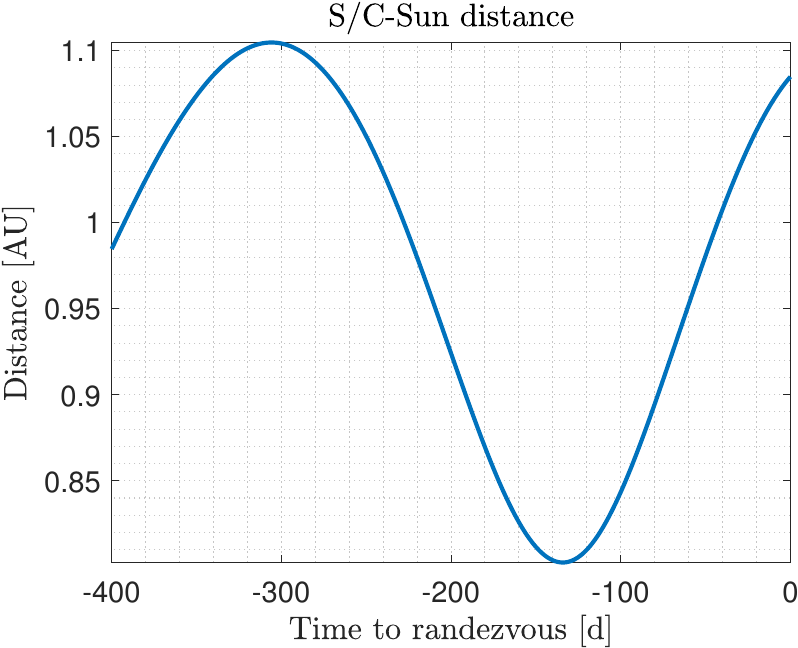}}
\subfigure[][{Target phase angle.}]
{\includegraphics[width=0.48 \textwidth]{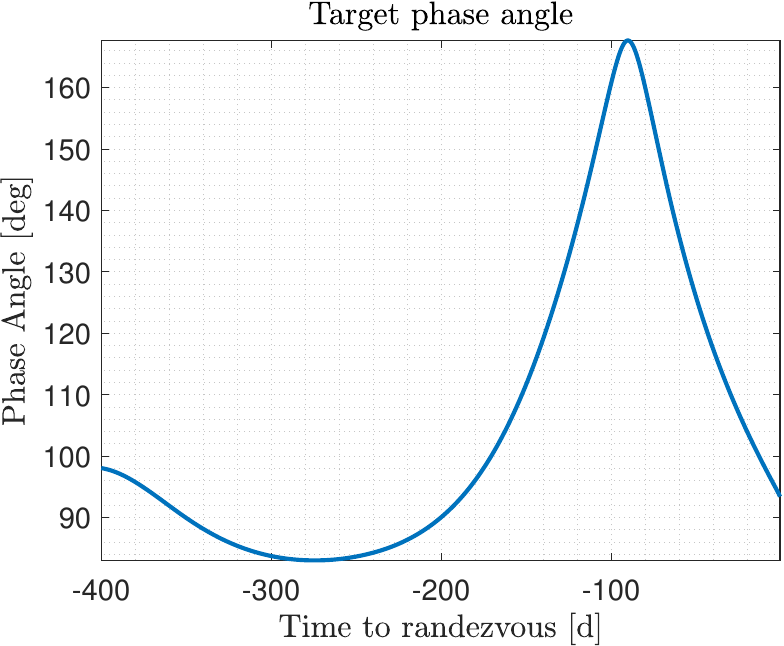}}\\
\subfigure[][{Trajectory representation.}]
{\includegraphics[width=0.7 \textwidth]{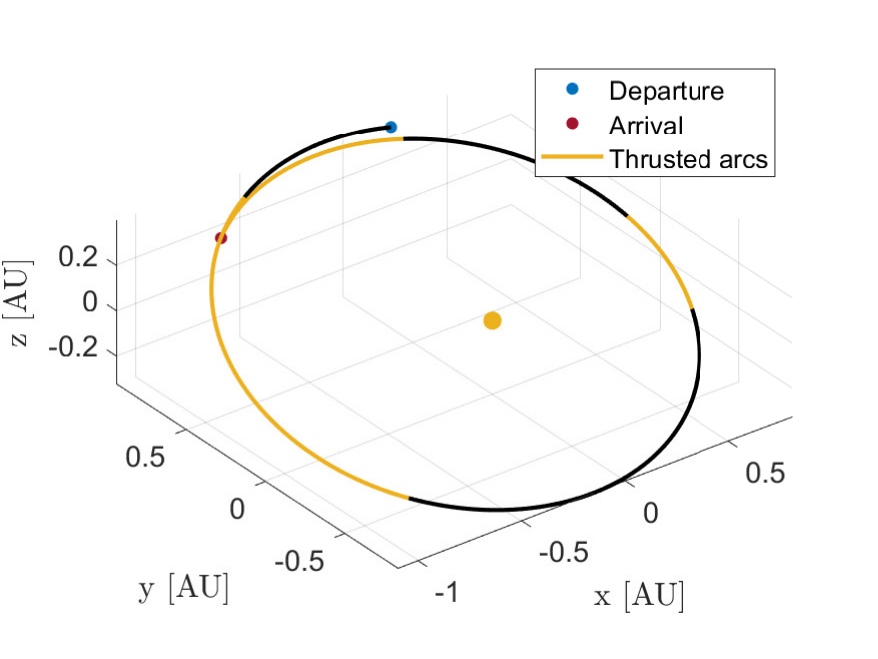}}
\caption{Geometrical analysis and representation in the J2000 reference frame of the solution C.}
\label{fig:geom_dec_traj}
\end{figure}

From the reachability analysis presented in Fig.~\ref{fig:porkchop}, three sample trajectories have been selected, indicated with three dots:

\begin{enumerate}[label=\Alph*.]
    \item (dark orange dot): Departure date on May 10, 2027;
    \item (yellow dot): Departure date on November 11, 2027;
    \item (white dot): Departure date on December 11, 2027.
\end{enumerate}
The three trajectories have been selected because they are respectively relevant due to
\begin{enumerate*}[label=\Alph*.]
    \item an early launch date that would allow backup solutions in case of failure,
    \item a convenient required propellant mass significantly lower than the others, and
    \item a later launch date that would still allow an early arrival at Apophis with required propellant mass similar to the solution A.
\end{enumerate*}
No trajectory has been selected in the first available window (i.e., between November and December 2026) as a launch in this period could be too early for the completion of the mission design phases. The characteristics of the selected trajectories have been reported in Table~\ref{tab:selected_trajectories}. Note that each of them requires a propellant mass value lower than \qty{73}{\kilogram}, and arrives two-to-three months before the close encounter to ensure sufficient time for pre-fly-by characterization of Apophis. The thrusting profiles of the selected trajectories are shown in Fig.~\ref{fig:selected_trajectories_profiles}. In Figs.~\ref{fig:geom_may_traj}--\ref{fig:geom_dec_traj} some preliminary geometrical considerations relative to the selected transfers are reported, as well as the interplanetary trajectories in the heliocentric equatorial frame.

\begin{table}[ht!]
    \centering
    \begin{tabular}{lccccc}
        \toprule
        \toprule
             \textbf{ID} & \textbf{Departure date} & \textbf{Arrival date} & \textbf{ToF} (d) &$\mathbf{m}_p$ (kg)& $\boldsymbol{\Delta} \mathbf{V}$ (km/s)\\
                \midrule

            A & May 10, 2027 & Feb 8, 2029 & 640 & 70.43 & 2.21\\
            B  & Nov 11, 2027 & Jan 23, 2029 & 439 & 54.95 & 1.69 \\
            C  & Dec 11, 2027 & Jan 14, 2029 & 400 & 72.23 & 2.29 \\
        \bottomrule
        \bottomrule
    \end{tabular}
    \caption{Characteristics of the selected representative trajectories.}
    \label{tab:selected_trajectories}
\end{table}

\noindent As expected, solution A exhibits a rather empty thrust profile, while trajectory C requires longer and more frequent thrusting arcs. This behaviour correlates with the distance from the time-optimal solution. On the other hand, there is no significant differences in the distance and in the geometric quantities. This finding can simplify the system design, since communications with the ground, power generation, and relative navigation with the target are not severely affected by the trajectory of choice.

\subsection{Analyses with Higher Thrust-to-Mass Ratios}
\begin{figure}[!ht]
\centering
\subfigure[][{$T_{\text{max}}/m_0 = 1.8 \times 10^{-4} m/s^2$.}]
{\includegraphics[width=0.48 \textwidth]{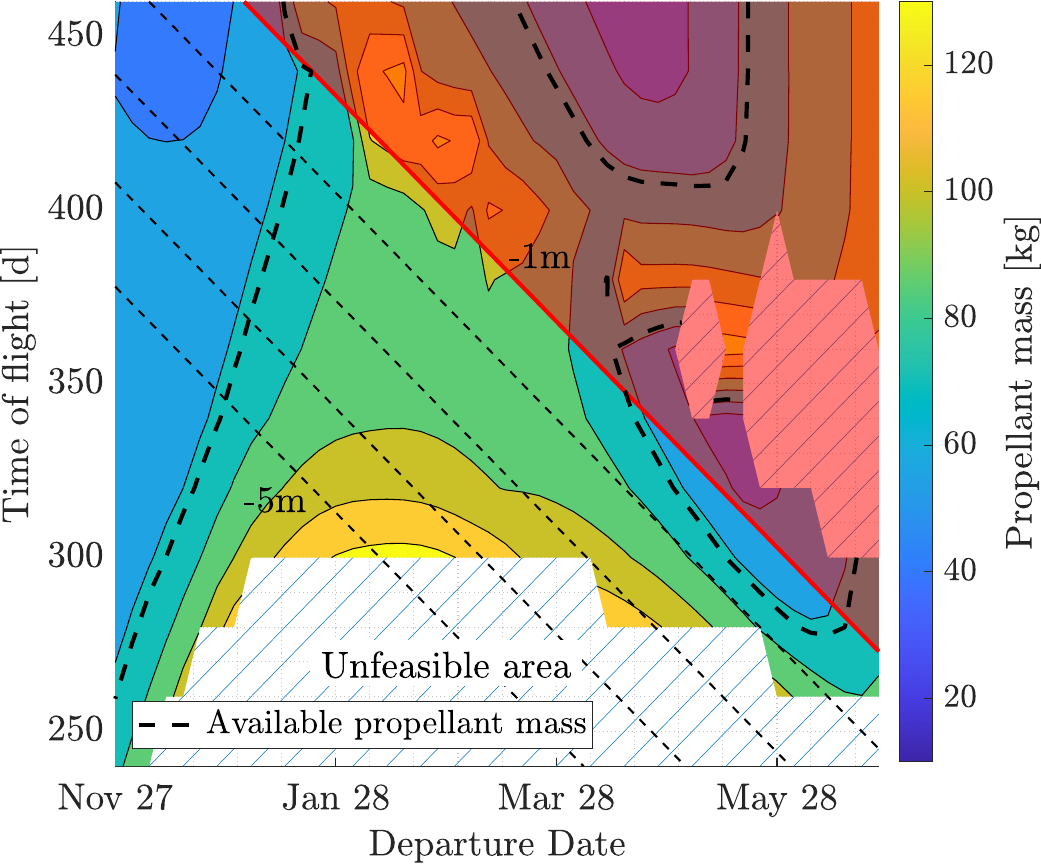}}
\subfigure[][{$T_{\text{max}}/m_0 = 2 \times 10^{-4} m/s^2$.}]
{\includegraphics[width=0.48 \textwidth]{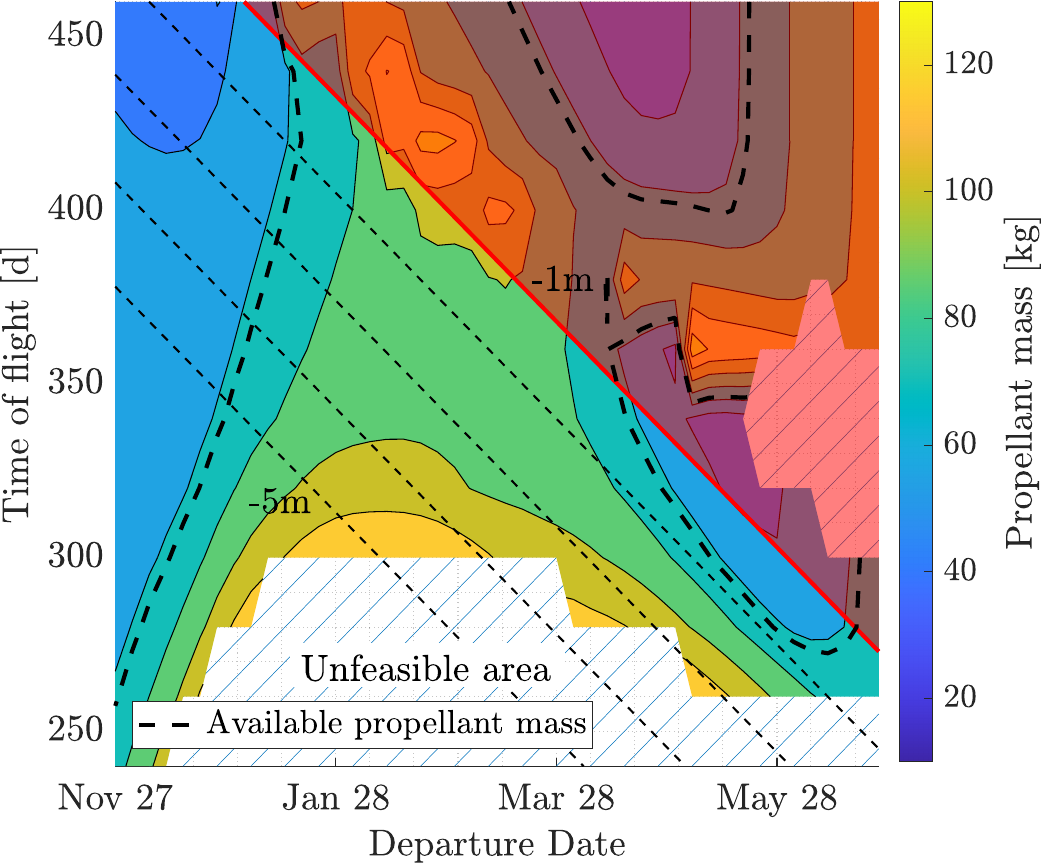}}
\caption{Porkchop plots for higher thrust-to-mass ratios.}
\label{fig:higherTM}
\end{figure}

The nominal analysis has been conducted considering a $T_{\text{max}}/m_0$ of \qty{1.2 e-4}{\meter\per\second\squared}, corresponding to a thrust of \qty{60}{\milli\newton}. In order to explore if additional opportunities are present when different propulsive systems are considered, a sensitivity analysis is performed, considering higher thrust-to-mass ratios, on the bottom right portion of Fig.~\ref{fig:porkchop}. A possible window in this region will allow a late launch to the asteroid, giving more time for the design and integration of the spacecraft.\\
Figs.~\ref{fig:higherTM} show that an additional window that satisfies the time and propellant constraints opens up in the timespan March--May 2028 with 270-360 days of time of flight, when thrusters with slightly higher thrust-to-mass ratios are considered (namely, \qty{1.8 e-4}{\meter\per\second\squared} and \qty{2 e-4}{\meter\per\second\squared} corresponding to, respectively, thrust levels of \qty{90}{\milli\newton} and \qty{100}{\milli\newton} and the nominal initial mass of \qty{500}{\kilogram}). For these plots, the patched red area indicates rendezvous happening later than one month before the close encounter of the asteroid with the Earth. In conclusion, higher thrust levels may open up new feasible launch windows and thus improve the asteroid reachability.

\section{Conclusion}
\label{sec:conclusions}
In the context of the RAMSES mission, a fast and robust methodology has been introduced to evaluate the reachability of Apophis with a low-thrust satellite before its close encounter with the Earth on April 13, 2029. To this aim, porckchop plots from the Earth to Apophis have been computed using a two-layer approach, exploiting a direct sequential convex programming algorithm followed by an indirect method. This method allows an easy handling of the free launching condition, while guaranteeing optimality of the solution, and it is able to compute more than 16,000 trajectories in few hours. Under the considered hypotheses, three feasible launching windows are identified. Each window lasts about 3 months and is separated from the other by about half a year, with the last one opening in September 2027. A fourth late short window, between April and May 2029, can be found if higher thrust-to-mass ratios are considered. 

\backmatter

\bmhead{Acknowledgments}

This work has been performed in response to ESA call AO/1-11659/23/NL/GLC: Pre-phase A/Phase A System Study of an Apophis Mission (RAMSES Definition Phase).

\bmhead{Conflict of interest}
The authors declare that they have no known competing financial interests or personal relationships that could have appeared to influence the work reported in this paper.

\bibliography{sn-bibliography}%

\end{document}